\documentclass[aps,10pt,prl,twocolumn,superscriptaddress]{revtex4-2}
\usepackage{graphicx}
\usepackage{amsmath}
\usepackage{amsfonts,hyperref}
\usepackage{amssymb,mathrsfs}
\usepackage{upgreek}
\usepackage{bbold,yfonts}

\newcommand{\bs}[1]{\boldsymbol{#1}}

\begin{document}

\title{Electronic viscosity and energy relaxation in neutral graphene}

\author{Vanessa Gall} \affiliation{\mbox{Institute for Quantum
    Materials and Technologies, Karlsruhe Institute of Technology,
    76021 Karlsruhe, Germany}} \affiliation{\mbox{Institut f\"ur
    Theorie der Kondensierten Materie, Karlsruhe Institute of
    Technology, 76128 Karlsruhe, Germany}}

\author{Boris N. Narozhny} \affiliation{\mbox{Institut f\"ur Theorie
    der Kondensierten Materie, Karlsruhe Institute of Technology,
    76128 Karlsruhe, Germany}} \affiliation{National Research Nuclear
  University MEPhI (Moscow Engineering Physics Institute), 115409
  Moscow, Russia}

\author{Igor V. Gornyi} \affiliation{\mbox{Institute for Quantum
    Materials and Technologies, Karlsruhe Institute of Technology,
    76021 Karlsruhe, Germany}} \affiliation{\mbox{Institut f\"ur
    Theorie der Kondensierten Materie, Karlsruhe Institute of
    Technology, 76128 Karlsruhe, Germany}} \affiliation{Ioffe
  Institute, 194021 St.~Petersburg, Russia}

\date{\today}

\begin{abstract}
  We explore hydrodynamics of Dirac fermions in neutral graphene in
  the Corbino geometry. In the absence of magnetic field, the bulk
  Ohmic charge flow and the hydrodynamic energy flow are decoupled.
  However, the energy flow does affect the overall resistance of the
  system through viscous dissipation and energy relaxation that has to
  be compensated by the work done by the current source. Solving the
  hydrodynamic equations, we find that local temperature and electric
  potential are discontinuous at the interfaces with the leads as well
  as the device resistance and argue that this makes Corbino geometry
  a feasible choice for an experimental observation of the Dirac
  fluid.
\end{abstract}

\maketitle

Quantum dynamics of charge carriers is one of the most important
research directions in condensed matter physics. In many materials
transport properties can be successfully described under the
assumption of weak electron-electron interaction allowing for
free-electron theories \cite{ziman}. An extension of this approach to
strongly-correlated systems remains a major unsolved problem. The
advent of ``ultra-clean'' materials poses new challenges, especially
if the electronic system is nondegenerate. At high temperatures such
systems may exhibit signatures of a collective motion of charge
carriers resembling the hydrodynamic flow of a viscous fluid
\cite{geim1,kim1,geim2,kim2,geim3,geim4,gal,imh,imm,young,mac,ihn,goo}.

\begin{figure}[t]
\centering
\includegraphics[width=0.4\columnwidth]{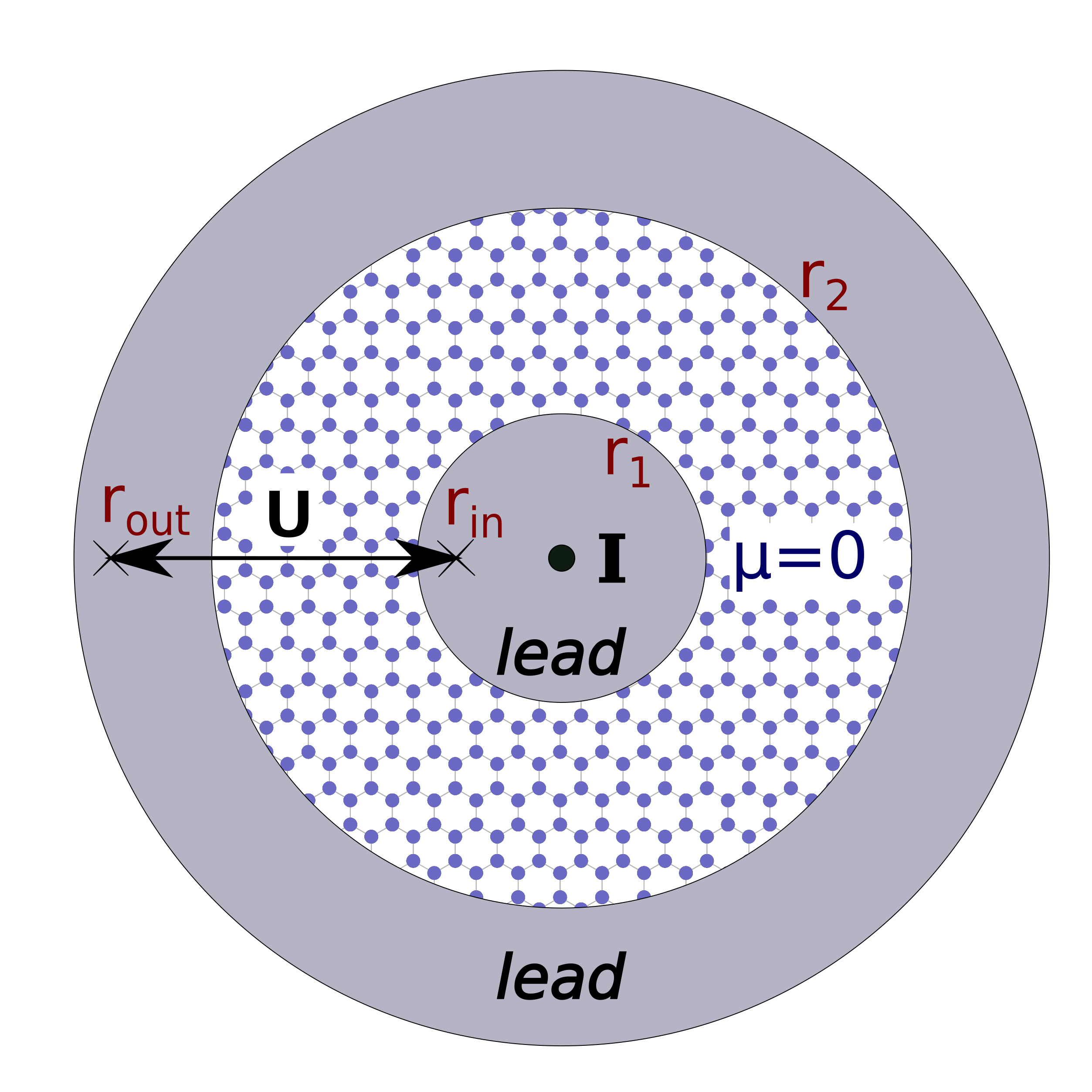}
\caption{Corbino geometry: the annulus-shaped sample of neutral
  graphene (${\mu=0}$) is placed between the the two leads: the inner
  circle of the radius $r_1$ and the outer shell with the inner radius
  $r_2$. A current $I$ is injected through at the center point and a
  voltage $U$ is measured between electrodes placed at the inner and
  outer radius $r_\text{in}$ and $r_\text{out}$.}
\label{fig:CorbinoGeneral}
\end{figure}

Electronic viscosity has been discussed theoretically for a long time
\cite{stein,gurzhi,read,ale,moo,me2}, but became the subject of
dedicated experiments \cite{geim1,imh} only recently, after
ultra-clean materials became available. Up until now, most
experimental efforts were focusing on graphene
\cite{geim1,kim1,geim2,kim2,geim3,geim4,gal,imh,imm,young} where the
hydrodynamic regime is apparently easier to achieve \cite{luc,rev}.
Viscous effects manifest themselves in nonuniform flows. In the common
``linear'' geometry (channels, wires, Hall bars, etc.) this occurs in
``narrow'' samples where the typical length scale associated with
viscosity is of the same order as the channel width
\cite{fl0,mr1,mr3,mr2,cfl}. In contrast, in the ``circular'' Corbino
geometry, see Fig.~\ref{fig:CorbinoGeneral}, the electric current is
nonuniform even in the simplest Drude picture (in the absence of
magnetic field, $\bs{j}\propto\bs{e}_{\bs r}/|\bs{r}|$, where
$\bs{e}_{\bs r}=\bs{r}/|\bs{r}|$) making it an excellent platform to
measure electronic viscosity \cite{corbino,corb,ady,fal19}. In the
last year, electronic hydrodynamics in the Corbino geometry has been
studied both experimentally \cite{sulp22} and theoretically
\cite{oga21,alex22,ady22,rai22}.

In this paper we address the ``Dirac fluid'' \cite{kim1,imh} (the
hydrodynamic flow of charge carriers in neutral graphene) in the
Corbino geometry. Unlike doped graphene where degenerate,
Fermi-liquid-like electrons may be described by the Navier-Stokes
equation with a weak damping term due to disorder
\cite{gurzhi,fl0,luc}, the two-band physics of neutral graphene leads
to unconventional hydrodynamics \cite{rev,me1}. In the hydrodynamic
approach any macroscopic current can be expressed as a product of the
corresponding density and hydrodynamic velocity $\bs{u}$ (up to
dissipative corrections), e.g., the electric and energy current
densities are ${\bs{j}=n\bs{u}}$ and ${\bs{j}_E=n_E\bs{u}}$,
respectively. In the degenerate regime the charge and energy densities
are proportional to each other (to the leading approximation in
thermal equilibrium ${n_E=2\mu n/3}$, where $\mu$ is the chemical
potential) and the two currents are equivalent \cite{hydro0}.  In
contrast, the equilibrium charge density vanishes at charge
neutrality, ${n(\mu=0)=0}$, while the energy density remains finite.
The two currents ``decouple'': the energy current remains
``hydrodynamic'', the charge current is completely determined by
the dissipative correction $\delta\bs{j}$.

Electronic transport at charge neutrality has been a subject of
intensive research
\cite{imh,mrexp,kash,mus,alf,schutt,hydro0,hydro1,julia,mr1,mr3,mr2,cfl,megt2}
leading to general consensus on the basic result: in the absence of
magnetic field, ${\bs{B}=0}$, resistivity of neutral graphene is
determined by the electron-electron interaction
\begin{equation}
\label{r0}
R_0 =
\frac{\pi}{2e^2T\ln2}\left(\frac{1}{\tau_{11}}+\frac{1}{\tau_{\rm
    dis}}\right) \underset{\tau_{\rm
    dis}\rightarrow\infty}{\longrightarrow} \frac{1}{\sigma_Q}.
\end{equation}
Here $\tau_{11}\propto\alpha_g^{-2}T^{-1}$ describes the appropriate
electron-electron collision integral and $\sigma_Q$ is the
``intrinsic'' or ``quantum'' conductivity of graphene. Disorder
scattering is characterized by the mean free time $\tau_{\rm dis}$,
which is large under the assumptions of the hydrodynamic regime,
$\tau_{\rm dis}\gg\tau_{11}$ and yields a negligible contribution to
Eq.~(\ref{r0}). Equation (\ref{r0}) describes the uniform bulk current
and is independent of viscosity (i.e., in a channel
\cite{hydro1,luc,mr1,megt2}). In contrast, in the Corbino geometry the
current flow is necessarily inhomogeneous and hence viscous
dissipation must be taken into account.

We envision the following experiment: a graphene sample (at charge
neutrality) in the shape of an annulus is placed between the inner (a
disk of radius $r_1$) and outer (a ring with the inner radius $r_2$)
metallic contacts (leads). For simplicity, we assume both leads to be
of the same material, e.g., highly doped graphene with the same doping
level. The electric current $I$ is injected into the center of the
inner lead preserving the rotational invariance (e.g., through a thin
vertical wire attached to the center point) and spreads towards the
outer lead, which for concreteness we assume to be grounded. The
overall voltage drop $U$ is measured between two points in the two
leads (at the radii ${r_{\rm in}<r_1}$ and ${r_{\rm{out}}>r_2}$)
yielding the device resistance, ${R=U/I}$. In most traditional
measurements, the leads' resistance is minimal, while the contact
resistance is important only in ballistic systems, see e.g.,
Ref.~\cite{imm}. Hence, one may interpret the measured voltage drop in
terms of resistivity of the sample material. Here we focus on the
device resistance and show that in the hydrodynamic regime there is an
additional contribution due to electronic viscosity and energy
relaxation.

\begin{figure}[t]
\centering
\includegraphics[width=0.8\columnwidth]{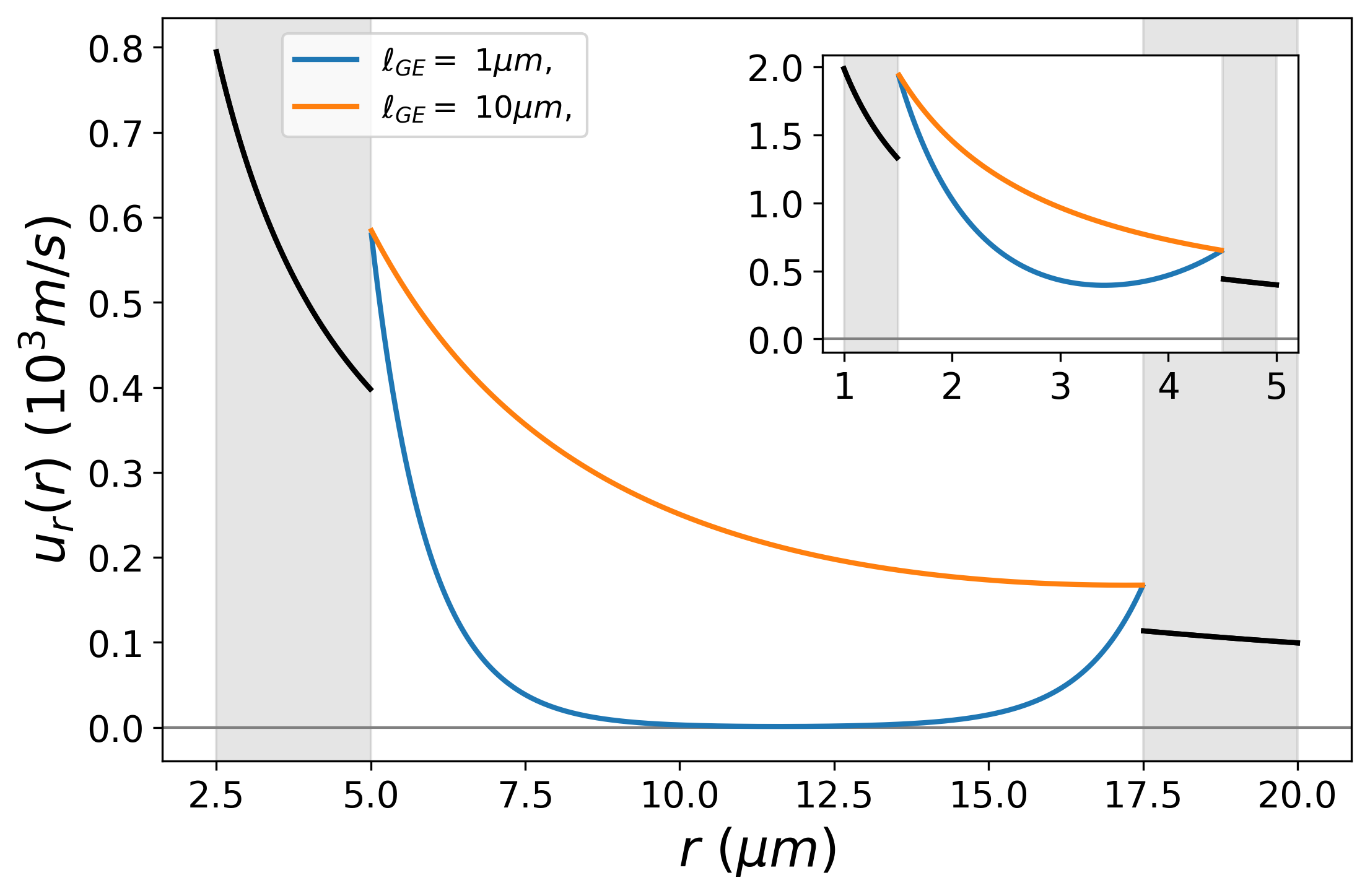}
\caption{Radial component of the hydrodynamic velocity $u_r$. Black
  lines show the drift velocity in the leads,
  $u_r^{\rm in(out)}\propto1/r$. Colored curves correspond to the solution
  Eq.~(\ref{u1}) for the two indicated values of $\ell_{GE}$. The
  results are plotted for the two cases of a large (main panel) and
  small (inset) device. }
\label{fig:ur}
\end{figure}

Charge flow through the Corbino disk can be described as follows. The
injected current spreads through the inner lead according to the Ohm's
law and continuity equation. In the stationary case, the latter
determines the radial component of the current density,
${j^{\rm in}_r=I/(2\pi{e}r)}$. This defines the drift velocity
${\bs{u}^{\rm in}=\bs{j}^{\rm in}/n^{\rm in}}$ ($n^{\rm in}$ is the carrier density in
the inner lead) and the energy current
${\bs{j}^{\rm in}_E=n_E^{\rm in}\bs{u}^{\rm in}}$. Reaching the interface, both
currents continue to flow into the graphene sample. Here (at $n=0$ and
${\bs{B}=0}$) the energy current ${\bs{j}_E=n_E\bs{u}}$ is decoupled
from the electric current ${\bs{j}=\delta\bs{j}}$. Charge conservation
requires the radial component of the electric current to be continuous
at the interface, ${\delta\bs{j}(r_1)=\bs{j}^{\rm in}(r_1)}$. Due to the
continuity equation, the current density in graphene has the same
functional form, ${\delta j_r=I/(2\pi{e}r)}$. Does this mean that the
device resistance trivially follows if one knows the resistivity of
graphene?  The answer is ``no'', since the electrochemical potential
is discontinuous at the interface! There are two mechanisms for the
``jump'' of the potential: (i) the usual Schottky contact resistance
\cite{mard,alf}, and (ii) dissipation due to viscosity \cite{fal19}
and energy relaxation \cite{meig1}. Since the lost energy must come
from the current source, both contribute to $R$.

The energy flow in neutral graphene is described by the set of
hydrodynamic equations developed in Refs.~\cite{hydro1,me1,meig1} and
most recently solved in Ref.~\cite{megt2} in the channel
geometry. Within linear response, the equations are
\begin{subequations}
\label{hydrolin1}
\begin{equation}
\label{cen2}
\bs{\nabla}\!\cdot\!\delta\bs{j} = 0,
\end{equation}
\begin{equation}
\label{ceni2}
n_{I}\bs{\nabla}\!\cdot\!\bs{u} + \bs{\nabla}\!\cdot\!\delta\bs{j}_I 
= - (12\ln2/\pi^2)n_{I}\mu_I/(T\tau_R),
\end{equation}
\begin{eqnarray}
\label{nseqlin1}
\bs{\nabla} \delta P
=
\eta \Delta\bs{u}
-
3P\bs{u}/(v_g^2\tau_{{\rm dis}}),
\end{eqnarray}
\begin{equation}
\label{tteqlin1}
3P\bs{\nabla}\!\cdot\!\bs{u} = -2\delta P/\tau_{RE}.
\end{equation}
\end{subequations}
Here Eq.~(\ref{cen2}) is the continuity equation; Eq.~(\ref{ceni2}) is
the ``imbalance'' continuity equation \cite{alf,me1} ($\mu_I$ is the
imbalance chemical potential, $n_{I}=\pi T^2/(3v_g^2)$ is the
equilibrium imbalance density, $v_g$ is the band velocity in graphene,
and $\tau_R$ is the recombination time); Eq.~(\ref{nseqlin1}) is the
linearized Navier-Stokes equation \cite{msf,me1,megt,megt2}; and
Eq.~(\ref{tteqlin1}) is the linearized ``thermal transport'' equation
($\tau_{RE}$ is the energy relaxation time \cite{meig1}). Equilibrium
thermodynamic quantities (the pressure $P=3\zeta(3)T^3/(\pi v_g^2)$,
enthalpy density ${\cal{W}}$, and energy density are related by the
``equation of state'', ${\cal{W}}=3{P}=3{n}_E/2$. The dissipative
corrections to the macroscopic currents are given by
\begin{subequations}
\label{djs}
\begin{equation}
\label{lj}
\delta\bs{j} = \bs{E}/(eR_0),
\end{equation}
\begin{equation}
\label{lji}
\delta\bs{j}_I = 
-\frac{2\gamma\ln2}{\pi}\, T\tau_{\rm dis}\bs{\nabla}\mu_I,
\quad
\gamma = \frac{\delta_I}{1\!+\!\tau_{\rm dis}/(\delta_I\tau_{22})},
\end{equation}
\end{subequations}
where $\tau_{22}\propto\alpha_g^{-2}T^{-1}$ describes a component of
the collision integral that is qualitatively similar, but
quantitatively distinct from $\tau_{11}$ and $\delta_I\approx 0.28$.
The equations (\ref{hydrolin1}) and (\ref{djs}) should be solved for
$\bs{u}$, $\delta\bs{j}$, $\delta\bs{j}_I$, $\bs{E}$, $\mu_I$, and
$\delta P$.

Excluding $\delta P$ from Eqs.~(\ref{nseqlin1}) and (\ref{tteqlin1})
we find a second-order differential equation for $\bs{u}$
\begin{subequations}
\label{u1}
\begin{equation}
\eta' \Delta\bs{u}
=
3P\bs{u}/(v_g^2\tau_{{\rm dis}}),
\quad
\eta'=\eta+3P\tau_{RE}/2.
\end{equation}
In the Corbino disk, the general solution for the radial component of
the velocity has the form
\begin{equation}
u_r = a_1 I_1\left(\frac{r}{\ell_{GE}}\right) + a_2 K_1\left(\frac{r}{\ell_{GE}}\right),
\quad
\ell_{GE}^2=\frac{v_g^2\eta'\tau_{\rm dis}}{3P},
\end{equation}
\end{subequations}
where $I_1(z)$ and $K_1(z)$ are the Bessel functions. The coefficients
$a_1$ and $a_2$ can be found using the continuity of the entropy
current at the two interfaces (within linear response). The resulting
behavior in shown in Fig.~\ref{fig:ur} (here we choose to show our
results in graphical form since the analytic expressions are somewhat
cumbersome \cite{suppl}; quantitative calculations were performed for
$T=100\,$K and experimentally relevant values of the parameters taken
from Refs.~\cite{imm,gal,imh,meig1}).

In the hydrodynamic regime, the electron-electron scattering time is
the shortest scale in the problem, hence the spatial variation of
$\bs{u}$ is determined by energy relaxation. If
${\ell_{GE}\ll{r_{\rm{out}}-r_{\rm in}}}$, 
then the energy current
injected from the leads decays in a (relatively small) boundary region
while in the bulk of the sample $\bs{u}\rightarrow0$. On the other
hand, if $\ell_{GE}$ is of the same order as (or larger than) the
system size, then $u_r$ does not vanish and approaches the standard
Corbino profile, $u_r\propto1/r$. At each interface, $u_r$ exhibits a
jump due to the mismatch of the entropy densities in the sample and
leads.

\begin{figure}[t]
\centering
\includegraphics[width=0.8\columnwidth]{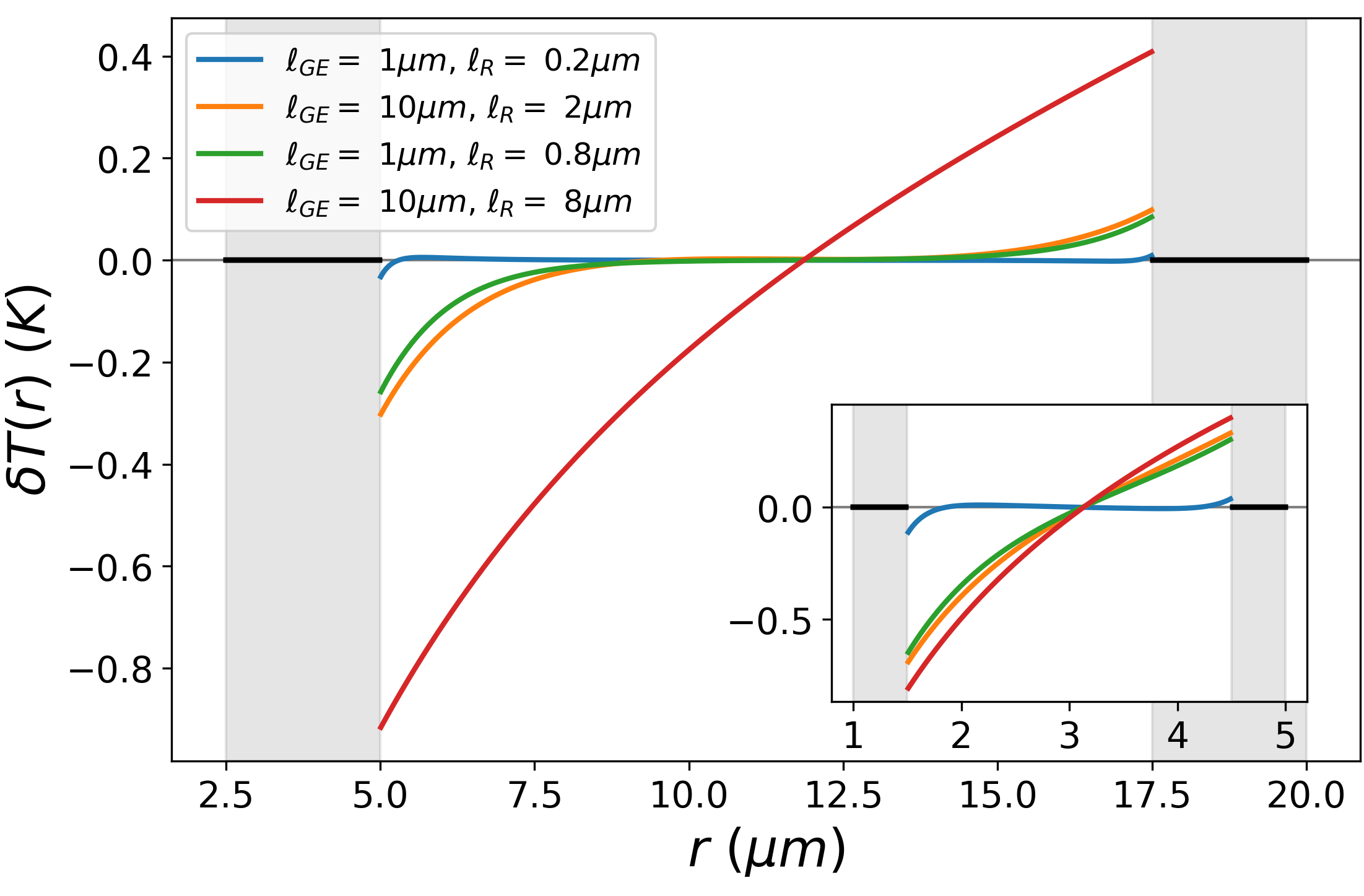}
\caption{Temperature distribution in the device. Colored curves
  correspond to the solution of the hydrodynamic equations for the
  indicated values of $\ell_{GE}$ and $\ell_{R}$. The results are
  plotted for the two cases of a large (main panel) and small (inset)
  device. In the leads $\delta T=0$, shown by black lines.}
\label{fig:t}
\end{figure}

The nonequilibrium quantities $\delta P$ and $\mu_I$ can now be found
straightforwardly. The former follows directly from
Eq.~(\ref{tteqlin1}) using the solution (\ref{u1}), while the
differential equation for the latter can be found by substituting
Eq.~(\ref{lji}) into Eq.~(\ref{ceni2}) and using the solution
(\ref{u1}). The boundary conditions for $\delta P$ and $\mu_I$ follow
from the continuity equations for the charge and imbalance. The two
quantities can be combined to determine the nonequilibrium temperature
variation, $\delta T$, shown in Fig.~\ref{fig:t}. For a large sample
(${\ell_{GE},\ell_R\ll{r_{\rm{out}}-r_{\rm in}}}$, 
$\ell_R^2=\gamma v_g^2\tau_{\rm dis}\tau_R/2$),
$\delta T$
exhibits fast decay and vanishes in the bulk of the sample. For larger
values of ${\ell_{GE},\ell_R}$ energy relaxation is less effective and
the system exhibits an inhomogeneous temperature profile.

The obtained solutions completely describe the hydrodynamic energy
flow in neutral graphene. Our remaining task is to find the behavior
of the electrochemical potential at the two interfaces enabling us to
determine $R$.

The standard description of interfaces between metals or
semiconductors \cite{mard} can be carried over to neutral graphene
\cite{alf} in terms of the contact resistance. Typically, this is a
manifestation of the difference of work functions of the two materials
across the interface. In graphene, the contact resistance was recently
measured in Ref.~\cite{imm}, see also
Refs.~\cite{sulp22,gall21,hak}. In the standard diffusive (or Ohmic)
case, the contact resistance leads to a voltage drop that is small
compared to the voltage drop in the bulk of the sample and can be
ignored. In contrast, in the ballistic case there is almost no voltage
drop in the bulk, such that most energy is dissipated at the contacts.
Both scenarios neglect electron-electron interactions.

In the diffusive case interactions lead to corrections to the bulk
resistivity \cite{aar,zna} and the contact resistance can still be
ignored. In the ballistic case electron-electron interaction may give
rise to a ``Knudsen-Poiseuille'' crossover \cite{gurzhi} and drive the
electronic system to the hydrodynamic regime. While the Ohmic
resistivity of the electronic fluid may remain small, the hydrodynamic
flow possesses another channel for dissipation through viscosity
\cite{fal19}. At charge neutrality, this effect is subtle, since the
electric current is decoupled from the hydrodynamic energy flow, see
Eq.~(\ref{lj}). At the same time, both are induced by the current
source providing the energy dissipated not only by Ohmic effects, but
also by viscosity \cite{fal19} and energy relaxation processes
\cite{meig1} that should be taken into account in the form of an
additional voltage drop. Since the voltage drop in the bulk of the
sample is completely determined by Eq.~(\ref{lj}), the additional
contribution takes the form of a jump in $\phi$ at the
interface corresponding to an excess electric
field induced in the thin Knudsen layer around the interface
\cite{fal19}.

\begin{figure}[t]
\centering
\includegraphics[width=0.8\columnwidth]{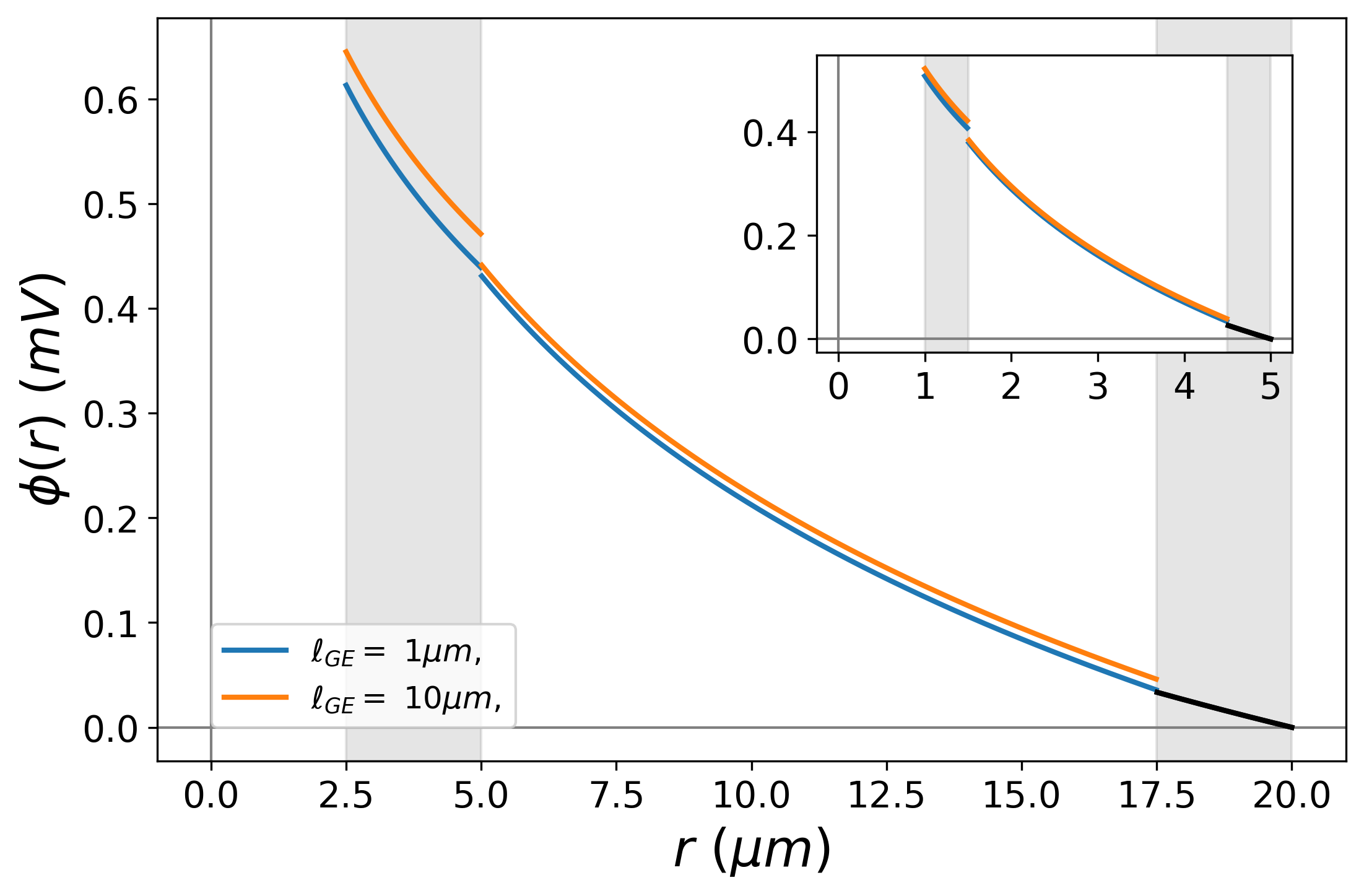}
\caption{Electrochemical potential (voltage drop) throughout the
  device. The black line shows the Ohmic behavior in the outer lead
  relative to the ground. The jumps at the interfaces are due to
  dissipative effects (viscosity and energy relaxation) in the bulk of
  the sample. }
\label{fig:phi}
\end{figure}

The magnitude of the jump in $\phi$ can be established by considering
the flow of energy through the interface. Following the standard route
\cite{fal19,dau6}, we consider the time derivative of the kinetic
energy, ${{\cal A}=\dot{\cal E}}$, where ${\cal E}$ is obtained by
integrating the energy density ${n_E(\bs{u})\!-\!n_E(0)}$ over the
volume. Working within linear response, we expand the latter to the
leading order in the hydrodynamic velocity. Finding time derivatives
from the equations of motion and using the continuity equation and
partial integration, we then separate the ``bulk'' and ``boundary''
contributions, ${\cal A}={\cal A}_{\rm bulk}+{\cal A}_{\rm edge}$. We
interpret the former as the bulk dissipation, while ${\cal A}_{\rm
  edge}$ includes the energy brought in (carried away) through the
boundary by the incoming (outgoing) flow. In the stationary state
$\dot{\cal E}=0$, dissipation is balanced by the work done by the
source. Assuming that no energy is accumulated at the interface, we
find the corresponding boundary condition.

\begin{figure}[t]
\centering
\includegraphics[width=0.8\columnwidth]{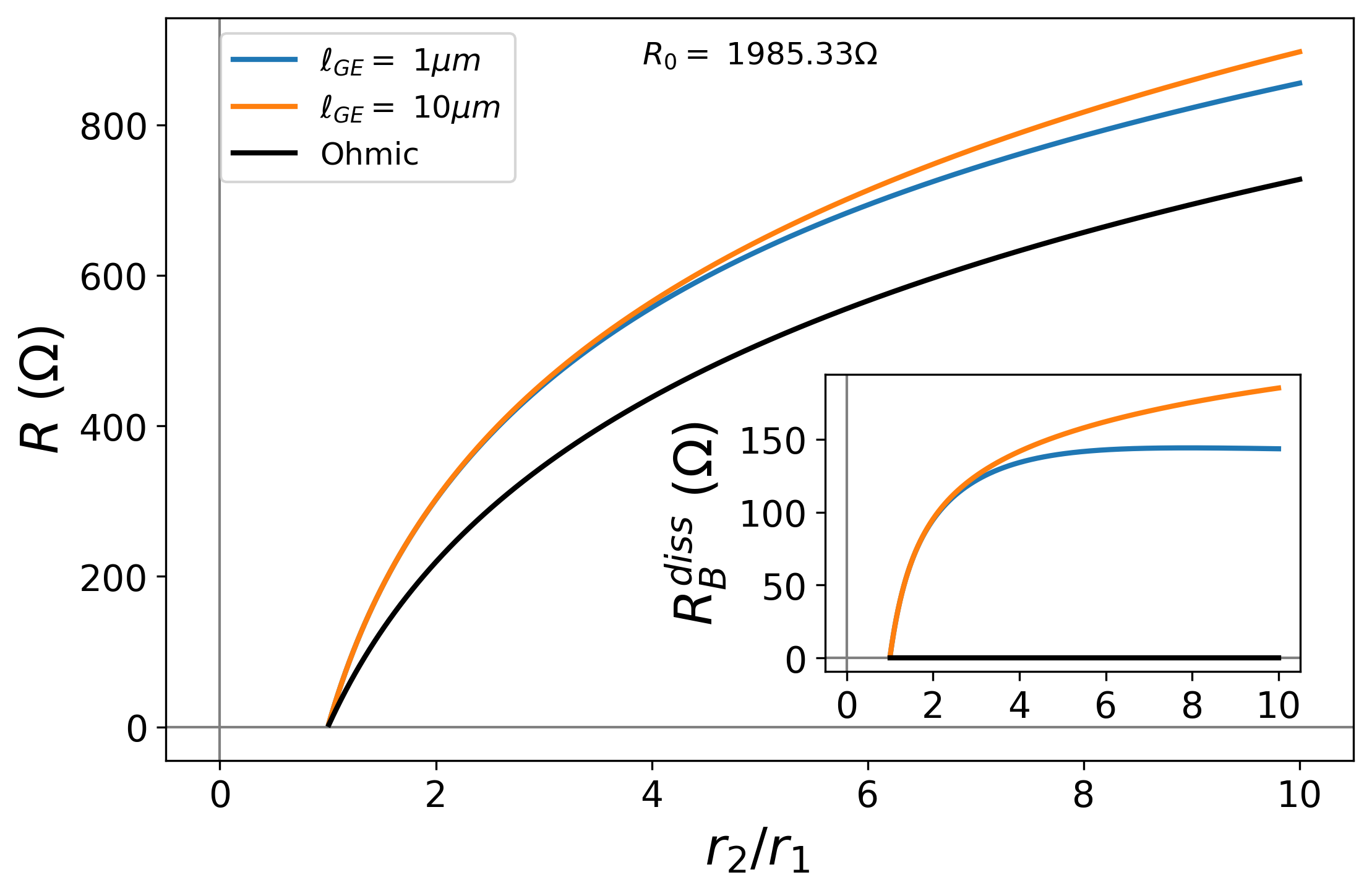}
\caption{Total resistance of the Corbino device for different values
  of $\ell_{GE}$ (here $r_1=0.5\,\mu$m). Inset: additional
  contribution to the resistance due to viscous dissipation.}
\label{fig:R}
\end{figure}

The specific form of the equations of motion depends on the choice of
the material. Assuming the leads' material is highly doped graphene,
the equation of motion is the usual Ohm's law with the diffusion term
\cite{df2} coming from the gradient of the stress-energy tensor
\cite{hydro0}, here we include a viscous contribution due to disorder
\cite{bur19} and find \cite{fal19} (omitting the continuous entropy
flux)
\begin{subequations}
\label{work}
\begin{equation}
\label{abl}
{\cal A}_{\rm edge}^{\rm lead} = \!\int\! dS_\beta 
\left(u^{L}_\alpha \sigma'_{L;\alpha\beta} - u^{L}_\beta \delta P_L - e \bs{j}^L_\beta \phi\right)\!,
\end{equation}
where $\bs{j}^L=n_L\bs{u}^{L}$ is the current density, $\bs{u}^{L}$ is
the drift velocity, $\delta P_L$ is the nonequilibrium pressure, and
$\sigma'_L$ is the viscous stress tensor in the lead. The first two
terms are the usual dissipative contributions to the energy flow
across the boundary \cite{dau6}, the last term is the Joule heat.

In neutral graphene, we obtain similar results from the
Navier-Stokes equation, except that the Joule heat is now determined
by $\delta\bs{j}$
\begin{equation}
\label{abs}
{\cal A}_{\rm edge}^{\rm sample} = \!\int\! dS_\beta 
\left(u_\alpha \sigma'_{\alpha\beta} - u_\beta \delta P - e\delta j_\beta \phi\right)\!.
\end{equation}
\end{subequations}
Equating the two contributions (\ref{work}) and using the above
solutions for the velocity and pressure, we find the jumps of the
potential $\phi$ at the two interfaces. This allows us to determine
$\phi$ everywhere in the device, see Fig.~\ref{fig:phi}, as well as
the device resistance.

The total resistance of the Corbino device is shown in
Fig.~\ref{fig:R}. Neglecting hydrodynamic effects, we find the usual
logarithmic dependence of $R$ on the system size. Viscosity and energy
relaxation provide an additional dissipation channel and hence
increase $R$. Energy relaxation contributes to this increase since it
dominates the hydrodynamic energy flow, see Eq.~(\ref{u1}). At the
same time, the boundary condition for the electric potential,
Eqs.~(\ref{work}), is determined by viscosity.

In this paper we have solved the hydrodynamic equations in neutral
graphene. We have shown, that despite the known decoupling of the
Ohmic charge flow and hydrodynamic energy flow, in Corbino geometry
the latter does affect the observable behavior leading to jumps in
temperature (shown in Fig.~\ref{fig:t}) and the electric potential,
see Fig.~\ref{fig:phi}. The potential jump is distinct from the usual
contact resistance insofar it is a function of the system size. Both
effects are observable using the modern imaging techniques (the local
temperature variation can be measured using the approach of
Refs.~\cite{halb,halb17,zel}, while measurements of the local
potential are at the heart of the technique proposed in
Refs.~\cite{sulp,imm}). Hydrodynamics also affects the more
conventional transport measurements through the size-dependent
contribution to the device resistance, see Fig.~\ref{fig:R}.

Our results highlight several particular features of the Dirac fluid
in neutral graphene. Firstly, the ``linear response'' currents
(\ref{djs}) are independent of the temperature gradient due to exact
particle-hole symmetry \cite{alf}. Secondly, in contrast to the case
of doped graphene \cite{fal19} {\it the Dirac fluid is compressible}
even within linear response (due to energy relaxation, see
Eq.~(\ref{tteqlin1}). Finally, the hydrodynamic flow in neutral
graphene is the energy flow. Hence, energy relaxation effectively
dominates over viscous effects, see Eqs.~(\ref{u1}), complicating
experimental determination of $\eta$.

External magnetic field is also known to couple the charge and energy
flows in neutral graphene \cite{me1}. We expect that our theory will
yield interesting results on Corbino magnetoresistance
\cite{gall21}. Another extension of our theory is the study of
thermoelectric phenomena, which is more interesting if one moves away
from the neutrality point \cite{alex22} (where the thermopower must
vanish due to the exact particle-hole symmetry). Our results on both
issues will be reported elsewhere.

\begin{acknowledgments}
 
The authors are grateful to P. Hakonen, V. Kachorovskii, A. Levchenko,
A. Mirlin, J. Schmalian, A. Shnirman, and M. Titov for fruitful
discussions. This work was supported by the German Research Foundation
DFG within FLAG-ERA Joint Transnational Call (Project GRANSPORT), by
the European Commission under the EU Horizon 2020 MSCA-RISE-2019
Program (Project 873028 HYDROTRONICS), by the German Research
Foundation DFG project NA 1114/5-1 (BNN), and by the German-Israeli
Foundation for Scientific Research and Development (GIF) Grant
No. I-1505-303.10/2019 (IVG).
\end{acknowledgments}

\onecolumngrid

\appendix

\section{Supplemental material}

Starting with the general form of the hydrodynamic equations in
graphene, we obtain the analytical results presented graphically in
the main text. In Sec.~\ref{sec:HydroGraphene} we summarize the
hydrodynamic equations for graphene. In Sec.~\ref{sec:HydroCorbino} we
specify these equations within linear response in polar coordinates at
charge neutrality and $B=0$. In Sec.~\ref{sec:hydroLeads} we formulate
a description of the leads followed by the relevant boundary
conditions at the lead-graphene interfaces in Sec.~\ref{sec:Boundary}.
Next, in Sec.~\ref{sec:fullSolution} we present the full analytical
solution for the hydrodynamic equations in the Corbino geometry with
the above boundary conditions. In Sec.~\ref{sec:Dissipation} we
discuss the dissipation in the system and corroborate the argument
used in the main text to obtain the device resistance. Finally, we
conclude with a brief analysis in Sec. \ref{sec:analysis}.

\subsection{Electronic hydrodynamics in graphene \label{sec:HydroGraphene}}

Following Ref.~\cite{Narozhny2021} we combine the chemical potentials
of the two bands in graphene $\mu_\pm$ into
\begin{align}
    \mu=(\mu_++\mu_-)/2,\quad \mu_I = (\mu_+-\mu_-)/2
\end{align}
and introduce their conjugate charge and imbalance densities
\begin{align}
    n= n_+-n_-,\quad n_I = n_++n_-.
\end{align}
Taking into account dissipative corrections due to electron-electron collisions we then obtain the electric $(\vec{j})$ and imbalance $(\vec{j}_I)$ currents as
\begin{align}
    \vec{j} = n\vec{u}+\delta\vec{j},\quad \vec{j}_I = n_I\vec{u}+\delta\vec{j}_I,
\end{align}
where $\vec{u}$ is the drift velocity. The energy current
$\vec{j}_E=n_E\vec{u}$ is proportional to the momentum density and is
not relaxed by electron-electron collisions.  The currents $\vec{j}$
and $\vec{j}_I$ satisfy the continuity equations
\begin{subequations}
\label{fulleqs}
\begin{equation}
\label{eq:charge}
\partial_t n + \vec{\nabla}\!\cdot\!\vec{j} = 0,
\end{equation}
which describes the exact conservation of charge and
\begin{equation}
\label{eq:imbalance}
\partial_t n_I + \vec{\nabla}\!\cdot\!\vec{j}_I = - \frac{n_I\!-\!n_{I,0}}{\tau_R}
=-\frac{12\ln2}{\pi^2}\frac{n_{I,0}\mu_I}{T\tau_R},
\end{equation}
where ${n_{I,0}=\pi{T}^2/(3v_g^2)}$ is the equilibrium value of the
total quasiparticle density (at $\mu_I=0$) and $\tau_R$ is the
recombination time. 

A similar equation can be formulated for the energy density
\begin{equation}
\label{eq:energy}
\partial_t n_E + \vec{\nabla}\!\cdot\!\vec{j}_E =e\vec{j}\vec{E}-\frac{n_E-n_{E,0}}{\tau_\text{RE}},
\end{equation}
where $\tau_\text{RE}$ is the energy relaxation time. Typically this
is replaced by the thermal transport equation
\begin{align}
&T\left[\frac{\partial s}{\partial t}
+
\vec{\nabla}_{\vec{r}}\!\cdot\!
\left(s\vec{u}-\delta\vec{j}\frac{\mu}{T}-\delta\vec{j}_I\frac{\mu_I}{T}\right)\right]
=
\delta\vec{j}\!\cdot\!
\left[e\vec{E}\!+\!\frac{e}{c}\vec{u}\!\times\!\vec{B}\!-\!T\vec{\nabla}\frac{\mu}{T}\right]
-
T\delta\vec{j}_I\!\cdot\!\vec{\nabla}\frac{\mu_I}{T}\nonumber\\
&+
\frac{\eta}{2}\left(\nabla_\alpha u_\beta \!+\! \nabla_\beta u_\alpha
\!-\! \delta_{\alpha\beta} \vec{\nabla}\!\cdot\!\vec{u}\right)^2-
\frac{n_E\!-\!n_{E,0}}{\tau_{RE}}
+
\mu_I \frac{n_I\!-\!n_{I,0}}{\tau_R}
+
\frac{{\cal W}\vec{u}^2}{v_g^2\tau_{\rm dis}}. \label{eq:thermalTransport}
\end{align}
Within linear response the two equations coincide.
Finally, the generalized Navier-Stokes equation is given by
\begin{align}
{\cal W}(\partial_t+\vec{u}\cdot\vec{\nabla})\vec{u}+v_g^2\vec{\nabla}P+\vec{u}\partial_tP+e(\vec{E}\cdot\vec{j})\vec{u}=v_g^2\left[\eta\Delta\vec{u}-\eta_H\Delta\vec{u}\times\vec{e}_B+en\vec{E}+\frac{e}{c}\vec{j}\times\vec{B}\right]-\frac{\vec{j}_E}{\tau_\text{dis}}. \label{eq:GeneralNS}
\end{align}
\end{subequations}
Here $\eta$ and $\eta_H$ are the shear and Hall viscosity coefficients, respectively.

The expressions for the dissipative corrections can be found in the
Appendix of Ref. \cite{Narozhny2021}.

\subsection{Charge neutral Corbino disk at $B=0$ \label{sec:HydroCorbino}}

Taking into account the rotational symmetry of the Corbino disk, we
express the hydrodynamic theory in polar coordinates
$(r,\vartheta)$. All quantities can only depend on the radial
component $r$. Within linear response and at $B=0$, the hydrodynamic
equations \eqref{fulleqs} can be transformed to
\begin{subequations}
\label{eq:polarCoordinates}
\begin{equation}
\label{6a}
\frac{1}{r}\frac{\partial(r\delta j_r)}{\partial r} = 0,
\end{equation}
\begin{equation}
\label{6b}
n_{I,0}\frac{1}{r}\frac{\partial(ru_r)}{\partial r} + \frac{1}{r}\frac{\partial(r \delta j_{Ir})}{\partial r}
= -\frac{12\ln2}{\pi^2}\frac{n_{I,0}\mu_I(r)}{T\tau_R},
\end{equation}
\begin{equation}
\label{6cang}
u_\vartheta=0,
\end{equation}
\begin{equation}
\label{6crad}
\frac{\partial \delta P}{\partial r}
=
\eta \partial_r\left(\frac{1}{r}\frac{\partial(ru_r)}{\partial r}\right)
-
\frac{3Pu_r}{v_g^2\tau_{{\rm dis}}},
\end{equation}
\begin{equation}
\label{6d}
3P\frac{1}{r}\frac{\partial(r u_r)}{\partial r} = - \frac{2\delta{P}(r)}{\tau_{RE}}.
\end{equation}
\end{subequations}
The electric field $\vec{E}$ does not appear in
Eqs.~(\ref{eq:polarCoordinates}) due to charge neutrality. It does
however determine the dissipative correction $\delta\vec{j}$ which at
charge neutrality is the whole current. In the absence of the magnetic
field, all currents are radial. In polar coordinates, the dissipative
corrections take the form
\begin{subequations}
\label{eq:polarDissipation}
\begin{equation}
\label{4arad}
\delta j_r = 
\frac{E_r(r)}
{e R_0},
\end{equation}
\begin{equation}
\label{4aang}
\delta j_\vartheta = 0,
\end{equation}
\begin{equation}
\label{4brad}
\delta j_{Ir} = 
-\frac{\delta_I}{\tau_{\rm dis}^{-1}\!+\!\delta_I^{-1}\tau_{22}^{-1}}
\frac{2T\ln2}{\pi}\frac{\partial \mu_I}{\partial r}\!,
\end{equation}
\begin{equation}
\label{4bang}
\delta j_{I\vartheta} = 0.
\end{equation}
\end{subequations}
Equations (\ref{eq:polarCoordinates}) and (\ref{eq:polarDissipation})
have to be solved together taking into account the corresponding
boundary conditions see below.

For the purposes of establishing the boundary conditions we also need
to specify the stress tensor. At $B=0$ (and within linear response,
meaning neglecting terms that are higher than the leading order in
velocity or its derivatives), the stress tensor is
\begin{equation}
\Pi^{\alpha\beta}_E = P\delta^{\alpha\beta} - \sigma^{\alpha\beta}.
\end{equation}
Since the Hall viscosity vanishes at charge neutrality (as well as at
$B=0$), the viscous stress tensor in Cartesian coordinates is given by
\begin{equation}
\sigma^{\alpha\beta} = \eta \left( \nabla^\alpha u^\beta + \nabla^\beta u^\alpha - 
\delta^{\alpha\beta} \vec{\nabla}\!\cdot\!\vec{u}\right),
\end{equation}
which in polar coordinates becomes
\begin{equation}
\sigma_{rr}=-\sigma_{\vartheta\vartheta}=\eta\left(\partial_r-\frac{1}{r}\right) u_r,
\quad \sigma_{r\vartheta}=\sigma_{\vartheta r}=\eta\left(\partial_r-\frac{1}{r}\right) u_\vartheta. 
\label{eq:defSigma}
\end{equation}

\subsection{Description of leads \label{sec:hydroLeads}}

The leads, which are attached at the inner and outer radius of the
Corbino disk, are assumed to be a normal metal in the degenerate
regime ($\mu_L\gg T$), where transport is dominated by disorder
scattering characterized by the relaxation time $\tau_L$. In this
case we may restrict ourselves to a single band, such that there is
a single macroscopic current satisfying the continuity equation
\begin{equation}
\partial_t n_L +\vec{\nabla}\vec{j}=0.
\end{equation}
Within linear response, one can obtain the macroscopic equation of
motion (or generalized Ohm's law) integrating the kinetic equation
\cite{Pellegrino2017}. This way one finds
\begin{equation}
m\partial_t \vec{j}+\vec{\nabla}\check{\Pi}_E-e n_L \vec{E}-\frac{e}{c}\vec{j}\times\vec{B}=-\frac{m}{\tau_L}\vec{j},
\end{equation}
where the stress tensor may me expressed in terms of thermodynamic
pressure and disorder-induced viscosity
\begin{equation}
{\Pi}_E^{\alpha\beta}=P\delta^{\alpha\beta} -{\sigma}^{\alpha\beta},
\qquad 
\eta_L=\frac{\mu^3\tau_L}{4 \pi v_g^2\hbar^2}.
\end{equation}

To be concrete, we assume that the leads' material is doped
graphene. In that case we may introduce the ``effective mass''
$m=\mu_L/v_g^2$ and the drift velocity $\vec{u}_L$, such that
$\vec{j}=n_L \vec{u}_L$. Expressing the carrier density in terms of
pressure, we find
\begin{equation}
m\vec{j}=\frac{3 P_L}{v_g^2}\vec{u}_L,
\end{equation}
where to lowest order in temperature we find $P_L=\mu^3/(3\pi
v_g^2\hbar^2)$.  In the stationary state and at $B=0$, the equations of motion become
\begin{eqnarray}
&&
\vec{\nabla}\vec{u}_L=0,
\\
&&
\nonumber\\
&&
\vec{\nabla}\check{\Pi}_E+e n_L  \vec{\nabla}\phi=-\frac{3 P_L}{v_g^2\tau_L}\vec{u}_L.
\end{eqnarray}
Experimentally, the density $n_L$ and the chemical potential $\mu$ are
fixed by the gate voltage. Moreover, under the common assumption of
fast equilibration in the leads, we may assume a uniform temperature
$T$ as well. The general variation of $P_L$ is found to be
\begin{align}
 \delta P_L=\left(\frac{2\pi\mu T\delta T}{3v_g^2}+\frac{\pi T^2\delta\mu}{3v_g^2}+\frac{\mu^2\delta\mu}{\pi v_g^2}\right)
\end{align}
and thus vanishes under the condition we consider. Since the leads are
highly doped, we find $n_L=n_+=n_I$, such that the imbalance chemical
potential $\mu_I$ vanishes.

\subsection{Boundary conditions \label{sec:Boundary}}

The differential equations \eqref{eq:polarCoordinates} and
\eqref{eq:polarDissipation} should be supplemented by a suitable set
of boundary conditions. The only boundaries present in the Corbino are
boundaries between the sample and the leads. Since charge conservation
is exact and also holds in the leads, we find
\begin{align}
j_r(r_1-\epsilon)=\delta j_r(r_1+\epsilon),\quad \delta j_r(r_2-\epsilon)=j_r(r_2+\epsilon).
\end{align}
Fixing the total current $I$ in a radially symmetric system completely
determines the current density
\begin{align}
\label{jr}
I=e\int \mathrm{d}\vec{A}\cdot\vec{j}=2\pi e r j_r.
\end{align}

In contrast, the total quasiparticle number (imbalance) and entropy
are not conserved due to recombination and energy relaxation
processes. However, assuming that the corresponding relaxation rates
are not singular at the interface, the continuity equations
(\ref{eq:imbalance}) and (\ref{eq:thermalTransport}) yield the
following boundary conditions for the radial components of the current
densities. 

The resulting boundary conditions at the two interfaces can be
summarized as follows

\begin{align}
j_r(r_1-\epsilon)=n_L u_r(r_1-\epsilon)&=\delta j_r(r_1+\epsilon) ,\\
j_{I,r}(r_1-\epsilon)=n_L u_{L,r}(r_1-\epsilon)&=n_{I,0}u_r(r_1+\epsilon)+\delta j_{I,r}(r_1+\epsilon)=\delta j_r(r_1+\epsilon) \label{eq:imbalance1},\\
s_L u_{L,r}(r_1-\epsilon)=s_B u_r(r_1+\epsilon) \label{eq:entropy1}
\end{align}
\begin{align}
j_r(r_2+\epsilon)=n_L u_{L,r}(r_2+\epsilon)&=\delta j_r(r_2-\epsilon),\\
j_{I,r}(r_2+\epsilon)=n_L u_{L,r}(r_2+\epsilon)&=n_{I,0}u_r(r_2-\epsilon)+\delta j_{I,r}(r_2-\epsilon)=\delta j_r(r_2-\epsilon)\label{eq:imbalance2},\\
s_L u_{L,r}(r_2+\epsilon)=s_B u_r(r_2-\epsilon)\label{eq:entropy2}.
\end{align}

\subsection{ Full solution \label{sec:fullSolution}}

Solving the equations of motion in the leads, we find
\begin{align}
u_{L,r}=\frac{I}{2\pi e n_L r},\quad u_{L,\vartheta}=0,\\
\sigma_{rr}=\frac{-I\eta_L}{\pi e n_L r^2},\quad \sigma_{r\vartheta}=0,\\
E_r = \frac{2 P_L}{e n_L v_g^2 \tau_L}\frac{I}{2\pi e n_L r},\\
\phi(r)=-\frac{I}{2\pi}\frac{2P_L}{e^2 n_L^2 v_g^2\tau_L}\log\left(\frac{r}{r_0}\right).
\end{align}
Here the drift velocity follows from the continuity equation and the
relation to the current which in turn is given by
Eq.~(\ref{jr}). After that, the assumption $\delta P=0$ leads to the
simple $1/r$ behavior for the electrical field $E_r$ as
well. Consequently, the charge density (from the Poisson equation) is
indeed constant. On the other hand, the constant $r_0$ in the
potential is not fixed by the boundary conditions we have imposed so
far. Finally, neither the electric field nor the current depend on the
disorder dominated viscosity $\eta_L$. However, the viscous stress
tensor itself is not zero, which will be used below later.

The above expressions can be re-written in terms of the temperature
$T$ and the chemical potential $\mu_L$ in the leads. Under our
assumptions, the leads' material is graphene, where the entropy
density is defined as
\begin{align}
Ts = 3P-\mu n-\mu_I n_I.
\end{align}
For $\mu\gg T$ in the leads we then find
\begin{align}
P_L&=\frac{\pi T^2\mu}{3v_g^2}+\frac{\mu^3}{3\pi v_g^2}=P_L^{T}+P_L^{T=0},\\
n_L&=\frac{\pi T^2}{3 v_g^2}+\frac{\mu^2}{\pi v_g^2}\\
s_L T&=3P_L-n_L\mu=\frac{\pi T^2\mu}{v_g^2}+\frac{\mu^3}{\pi v_g^2}-\frac{\pi T^2\mu}{3 v_g^2}-\frac{\mu^3}{\pi v_g^2}=\frac{2}{3}\frac{\pi T^2\mu}{v_g^2}=2 P_L^{T},
\end{align}
so we need to keep finite temperature corrections in the leads as well.

In our sample, the situation is more involved since in neutral
graphene the electric current is not related to the hydrodynamic
velocity. As a manifestation of this fact, the differential equations
\eqref{eq:polarCoordinates} and \eqref{eq:polarDissipation} decouple
into two disjunct sets. The first one consists of equations \eqref{6a}
and\eqref{4arad} with the solution
\begin{align}
\delta j_r& =\frac{I}{2\pi e r},\\
E_r &=\frac{I R_0}{2\pi r},\quad \phi=-\frac{I R_0}{2\pi }\log\left(\frac{r}{r_0}\right).
\end{align}
The constant $r_0$ (not necessarily the same as in the corresponding
solution for the leads) is not fixed by the boundary conditions we
have imposed so far.

The second set of equations consists of \eqref{6b}, \eqref{6crad},
\eqref{6d} and \eqref{4brad}. Expressing $\delta P$ through $u_r$, we
find
\begin{align}
 0 &= \partial_r\left(\frac{1}{r}\frac{\partial(r u_r)}{\partial r}\right)-\frac{u_r}{\ell_\text{GE}^2} \\
\frac{1}{\ell_\text{GE}^2}&=\left(\eta+\frac{3 P\tau_{RE} }{2}\right)^{-1}\frac{3P}{v_g^2\tau_{{\rm dis}}},
\end{align}
where the Gurzhi length is renormalized by energy relaxation through
the combination $\eta^\prime=\eta+3 P\tau_{RE}/2$. The other two
equations can be combined to form
\begin{align}
&\partial_r\left(\frac{1}{r}\frac{\partial(ru_r)}{\partial r}\right)-M\partial_r\left(\frac{1}{r}\frac{\partial(r \frac{\partial \mu_I}{\partial r})}{\partial r}\right)=-\frac{M}{\ell_R^2}\frac{\partial\mu_I(r)}{\partial r}\\
&M=\frac{2 T \ln 2}{n_{I,0}\pi}\frac{\delta_I}{\tau_{\rm dis}^{-1}\!+\!\delta_I^{-1}\tau_{22}^{-1}},\quad \ell_R^2=\frac{\delta_I}{\tau_{\rm dis}^{-1}\!+\!\delta_I^{-1}\tau_{22}^{-1}}\frac{\pi T^2\tau_R}{6 n_{I,0}}.
\end{align}
The two coupled Bessel differential equations for $u_r$ and
$\partial_r \mu_I $ can be expressed using the differential operator
$\mathbb{D}=\partial_r (1/r)\partial_r r$. This way we can write the
system of equations in the matrix form
\begin{align}
\mathbb{D}\begin{pmatrix}
1 & 0\\
1 & -M
\end{pmatrix}\begin{pmatrix}
u_r\\
\frac{\partial \mu_I}{\partial r}
\end{pmatrix}=\begin{pmatrix}
\frac{1}{\ell_\text{GE}^2} & 0\\
0 & -\frac{M}{\ell_R^2}
\end{pmatrix}\begin{pmatrix}
u_r\\
\frac{\partial \mu_I}{\partial r}
\end{pmatrix}\Leftrightarrow\mathbb{D}\begin{pmatrix}
u_r\\
\frac{\partial \mu_I}{\partial r}
\end{pmatrix}=\begin{pmatrix}
\frac{1}{\ell_\text{GE}^2}& 0\\
\frac{1}{M\ell_\text{GE}^2} & \frac{1}{\ell_R^2}
\end{pmatrix}\begin{pmatrix}
u_r\\
\frac{\partial \mu_I}{\partial r}
\end{pmatrix}.
\end{align}
This can be formally solved by diagonalizing the matrix
\begin{align}
\begin{pmatrix}
\frac{1}{\ell_\text{GE}^2}&0\\
\frac{1}{M\ell_\text{GE}^2} & \frac{1}{\ell_R^2}
\end{pmatrix}=\hat{U}^{-1}\hat{D}\hat{U},
\end{align} 
where $\hat{D}$ is a diagonal matrix with the eigenvalues $d_1$ and
$d_2$ (in units of inverse length squared) and then transforming
back to the $u_r$ and $\partial_r \mu_I $ basis. Then this coupled
Bessel differential equation has the general solution
\begin{align}
u_r&=M\left(1-\frac{\ell_\text{GE}^2}{\ell_R^2}\right)\left[f_1 I_1\left(\frac{r}{\ell_\text{GE}}\right)+f_2 K_1\left(\frac{r}{\ell_\text{GE}}\right)\right]\\
\frac{\partial\mu_I}{\partial r}&=f_1 I_1\left(\frac{r}{\ell_\text{GE}}\right)+f_2 K_1\left(\frac{r}{\ell_\text{GE}}\right)+g_1 I_1\left(\frac{r}{\ell_R}\right)+g_2 K_1\left(\frac{r}{\ell_R}\right),
\end{align}
where the coefficients $f_1$, $f_2$, $g_1$ and $g_2$ should be
determined from the boundary conditions. These involve the entropy density
\begin{align}
Ts_B = 3 P = 3\frac{3T^3\zeta(3)}{\pi v_g^2}.
\end{align}
From the conservation of entropy current Eqs.~\eqref{eq:entropy1} and
\eqref{eq:entropy2} we find $f_1$ and $f_2$ so that
\begin{align}
u_r=\frac{I s_L }{2 \pi  e n_L s_B}
\left\lbrace
\frac{I_1\left(\frac{r}{\ell_\text{GE}}\right) \left[r_1  K_1\left(\frac{r_1}{\ell_\text{GE}}\right)- r_2  K_1\left(\frac{r_2}{\ell_\text{GE}}\right)\right]
-K_1\left(\frac{r}{\ell_\text{GE}}\right) \left[ r_1  I_1\left(\frac{r_1}{\ell_\text{GE}}\right)- r_2  I_1\left(\frac{r_2}{\ell_\text{GE}}\right)\right]}
{ r_1 r_2 K_1\left(\frac{r_1}{\ell_\text{GE}}\right) I_1\left(\frac{r_2}{\ell_\text{GE}}\right)- r_1 r_2 I_1\left(\frac{r_1}{\ell_\text{GE}}\right) K_1\left(\frac{r_2}{\ell_\text{GE}}\right)}
\right\rbrace.
\end{align}
This leads to the stress tensor elements
\begin{align}
\sigma_{rr}&=\frac{\eta  I s_L}{2 \pi  e \ell_\text{GE} n_L s_B}
\frac{ I_2\left(\frac{r}{\ell_\text{GE}}\right) \left[r_1 K_1\left(\frac{r_1}{\ell_\text{GE}}\right)-r_2 K_1\left(\frac{r_2}{\ell_\text{GE}}\right)\right]+K_2\left(\frac{r}{\ell_\text{GE}}\right) 
\left[r_1 I_1\left(\frac{r_1}{\ell_\text{GE}}\right)-r_2 I_1\left(\frac{r_2}{\ell_\text{GE}}\right)\right]}
{ r_1 r_2  \left[K_1\left(\frac{r_1}{\ell_\text{GE}}\right) I_1\left(\frac{r_2}{\ell_\text{GE}}\right)-I_1\left(\frac{r_1}{\ell_\text{GE}}\right) K_1\left(\frac{r_2}{\ell_\text{GE}}\right)\right]},\\
\sigma_{r\vartheta}&=0
\end{align}
and
\begin{eqnarray}
&&
\delta P=-\frac{3 P \tau_\text{RE}}{2}\frac{1}{r}\frac{\partial (r u_r)}{\partial r}=
\\
&&
\nonumber\\
&&
\qquad
=
-\frac{3 P \tau_\text{RE}}{2}\frac{I  s_L}{2 \pi  e \ell_\text{GE} n_L s_B}
\left[\frac{K_0\left(\frac{r}{\ell_\text{GE}}\right) \left[ r_1 I_1\left(\frac{r_1}{\ell_\text{GE}}\right)- r_2  I_1\left(\frac{r_2}{\ell_\text{GE}}\right)\right]+I_0\left(\frac{r}{\ell_\text{GE}}\right) \left[ r_1  K_1\left(\frac{r_1}{\ell_\text{GE}}\right)-r_2 K_1\left(\frac{r_2}{\ell_\text{GE}}\right)\right]}{r_1 r_2  K_1\left(\frac{r_1}{\ell_\text{GE}}\right) I_1\left(\frac{r_2}{\ell_\text{GE}}\right)- r_1 r_2  I_1\left(\frac{r_1}{\ell_\text{GE}}\right) K_1\left(\frac{r_2}{\ell_\text{GE}}\right)}\right],
\nonumber
\end{eqnarray}
Using the conservation of the imbalance current
Eqs. \eqref{eq:imbalance1} and \eqref{eq:imbalance2} we find
the imbalance chemical potential
\begin{align}
\mu_I(r)&=\frac{I s_L\ell_R}{2 \pi  e M n_L r_1 r_2 s_B (\ell_\text{GE}^2-\ell_R^2) }
\left[
\frac{ K_0\left(\frac{r}{\ell_R}\right) \left[r_1 I_1\left(\frac{r_1}{\ell_R}\right)-r_2 I_1\left(\frac{r_2}{\ell_R}\right)\right] \left[\ell_\text{GE}^2 +\left(\ell_R^2-\ell_\text{GE}^2\right) \frac{n_L}{ n_{I,0}} \frac{s_B}{s_L}\right]}{ K_1\left(\frac{r_1}{\ell_R}\right) I_1\left(\frac{r_2}{\ell_R}\right)-I_1\left(\frac{r_1}{\ell_R}\right) K_1\left(\frac{r_2}{\ell_R}\right)}\right.
\nonumber\\
&\left.+\frac{ I_0\left(\frac{r}{\ell_R}\right) \left[r_1 K_1\left(\frac{r_1}{\ell_R}\right)-r_2 K_1\left(\frac{r_2}{\ell_R}\right)\right] \left[\ell_\text{GE}^2 +\left(\ell_R^2-\ell_\text{GE}^2\right) \frac{n_L}{ n_{I,0}} \frac{s_B}{s_L}\right]}{  K_1\left(\frac{r_1}{\ell_R}\right) I_1\left(\frac{r_2}{\ell_R}\right)-I_1\left(\frac{r_1}{\ell_R}\right) K_1\left(\frac{r_2}{\ell_R}\right)}\right.
\\
&\left.+\frac{ \ell_\text{GE} \ell_R  K_0\left(\frac{r}{\ell_\text{GE}}\right) \left[r_2 I_1\left(\frac{r_2}{\ell_\text{GE}}\right)-r_1 I_1\left(\frac{r_1}{\ell_\text{GE}}\right)\right]}{K_1\left(\frac{r_1}{\ell_\text{GE}}\right) I_1\left(\frac{r_2}{\ell_\text{GE}}\right)-I_1\left(\frac{r_1}{\ell_\text{GE}}\right) K_1\left(\frac{r_2}{\ell_\text{GE}}\right)} 
+
\frac{ \ell_\text{GE} \ell_R  I_0\left(\frac{r}{\ell_\text{GE}}\right) \left[r_2 K_1\left(\frac{r_2}{\ell_\text{GE}}\right)-r_1 K_1\left(\frac{r_1}{\ell_\text{GE}}\right)\right]}{K_1\left(\frac{r_1}{\ell_\text{GE}}\right) I_1\left(\frac{r_2}{\ell_\text{GE}}\right)-I_1\left(\frac{r_1}{\ell_\text{GE}}\right) K_1\left(\frac{r_2}{\ell_\text{GE}}\right)}
\right]
\nonumber
\end{align}
and the dissipative correction to the imbalance current
\begin{align}
\delta j_{Ir}(r)&=\frac{I n_{I,0} s_L}{2 \pi  e n_L r_1 r_2 s_B\left(\ell_\text{GE}^2-\ell_R^2\right)}
\left[
\frac{K_1\left(\frac{r}{\ell_R}\right) \left[r_1 I_1\left(\frac{r_1}{\ell_R}\right)-r_2 I_1\left(\frac{r_2}{\ell_R}\right)\right] \left[\ell_\text{GE}^2 +\left(\ell_R^2-\ell_\text{GE}^2\right) \frac{n_L}{ n_{I,0}} \frac{s_B}{s_L}\right]}{ K_1\left(\frac{r_1}{\ell_R}\right) I_1\left(\frac{r_2}{\ell_R}\right)-I_1\left(\frac{r_1}{\ell_R}\right) K_1\left(\frac{r_2}{\ell_R}\right)}\right.
\nonumber\\
&\left. 
-\frac{I_1\left(\frac{r}{\ell_R}\right) \left[r_1 K_1\left(\frac{r_1}{\ell_R}\right)-r_2 K_1\left(\frac{r_2}{\ell_R}\right)\right] \left[\ell_\text{GE}^2 +\left(\ell_R^2-\ell_\text{GE}^2 \right)\frac{n_L}{ n_{I,0}} \frac{s_B}{s_L}\right]}{K_1\left(\frac{r_1}{\ell_R}\right) I_1\left(\frac{r_2}{\ell_R}\right)-I_1\left(\frac{r_1}{\ell_R}\right) K_1\left(\frac{r_2}{\ell_R}\right)}
\right.
\\
&\left.
+\frac{\ell_R^2  K_1\left(\frac{r}{\ell_\text{GE}}\right) \left[r_1 I_1\left(\frac{r_1}{\ell_\text{GE}}\right)-r_2 I_1\left(\frac{r_2}{\ell_\text{GE}}\right)\right]}{K_1\left(\frac{r_1}{\ell_\text{GE}}\right) I_1\left(\frac{r_2}{\ell_\text{GE}}\right)-I_1\left(\frac{r_1}{\ell_\text{GE}}\right) K_1\left(\frac{r_2}{\ell_\text{GE}}\right)} 
+
\frac{\ell_R^2  I_1\left(\frac{r}{\ell_\text{GE}}\right) \left[r_2 K_1\left(\frac{r_2}{\ell_\text{GE}}\right)-r_1 K_1\left(\frac{r_1}{\ell_\text{GE}}\right)\right]}{K_1\left(\frac{r_1}{\ell_\text{GE}}\right) I_1\left(\frac{r_2}{\ell_\text{GE}}\right)-I_1\left(\frac{r_1}{\ell_\text{GE}}\right) K_1\left(\frac{r_2}{\ell_\text{GE}}\right)}
\right].\nonumber
\end{align}
From $\delta P$ and $\mu_I$ we find $\delta T$ according to
\begin{align}
\delta T = \frac{\pi v_g^2}{9 T^2\zeta(3)}\delta P-\frac{\pi^2}{27 \zeta(3)}\mu_I.
\end{align}

Our hydrodynamic system is not characterized by a local thermal
conductivity $\kappa$. In other words, the heat current
\begin{align}
\vec{j}_Q(r)=3 P\vec{u}-\mu\vec{j}-\mu_I\vec{j}_I
\end{align}
is related to the temperature gradient $\nabla \delta T(r)$ at the
same point $r$ non locally. The non-local (integral) relation between
$\vec{j}_Q(r)$ and $\nabla \delta T(r^\prime)$ characterized by a
non-local kernel $\kappa(r,r^\prime)$ follows from the fact that the
equation for $\vec{u}(r)$ is now a second-order differential equation
with a non-local Green's function. Expressing $\delta P(r)$ there in
terms of $\delta T(r)$ and $\mu_I(r)$, we have a non-local relation
between $\vec{u}(r)$, $\delta T(r^\prime)$ and $\nabla
\mu_I(r^\prime)$. Substituting this $\vec{u}(r)$ into the definition
of $\vec{j}_Q(r)$, we obtain a non-local thermal conductivity. As a
result one can only introduce the thermal conductance for the device,
relating the temperature difference between the contacts with the
total heat current through the system. This will be done in a
subsequent publication.

\subsection{Dissipation and total resistance \label{sec:Dissipation}}

The above solution is not sufficient to determine the drop in electrochemical potential between the
points $r_\text{in}$ and $r_\text{out}$ (in the inner and outer lead,
respectively) since it contains the undefined constant $r_0$ that has
to be determined from a boundary condition for the electric
potential. Although microscopically the potential has to be
continuous, several effects might contribute to an apparent
discontinuity on the hydrodynamic scale. The most obvious contribution
is the contact resistance that is a manifestation of the different
work functions in the two materials across the interface as well as
the mismatch in their chemical potentials \cite{Kamada2021b}. A more
subtle effect due to electron-electron interaction giving rise to
viscosity and hence an additional dissipation channel
\cite{Shavit2019}. At charge neutrality, this effect is subtle, since
the electric current is decoupled from the hydrodynamic energy flow.
However, both flows are induced by the same current source providing
the energy dissipated by means of both the Ohmic and viscous effects
\cite{Shavit2019} as well as energy relaxation \cite{meig1}. The
latter processes should be taken into account in the form of an
additional voltage drop at the interface.

Under the assumption that energy is not being accumulated at the
interface, we generalize the idea proposed in Ref.~\cite{Shavit2019}
and consider viscous dissipation in the sample. Since the electric
field in bulk of the sample is completely determined by the Ohmic
resistance $R_0$, additional dissipation due to viscosity and energy
relaxation corresponds to a jump in the electric potential (on the
hydrodynamic scale) at the interface. Microscopically, the voltage
jump is due to an excess electric field in the Knudsen layer around
the sample-lead boundary.

Consider the kinetic energy associated with the hydrodynamic flow that
can be found from the energy density
\begin{align}
\mathcal{E}= \int\mathrm{d}V \left(n_E-n_E(\vec{u}=0)\right)\approx \int\mathrm{d}V \frac{6P}{v_g^2}\vec{u}^2.
\end{align}
Working within linear response, here we only keep terms up to the second
order in $\vec{u}$ (and thus the drive $I$). Dissipation is then describe by 
\begin{align}
\mathcal{A}=\dot{\mathcal{E}}=2\frac{6P}{v_g^2}\int \mathrm{d V} \vec{u}\partial_t\vec{u}=0,
\end{align}
vanishing in the steady state. This expression can now be simplified
using the generalized Navier-Stokes equation.

In the leads (still assuming graphene at finite carrier density) we
find
\begin{align}
 \frac{3P_L}{v_g^2}\vec{u}_L\partial_t\vec{u}_L&=\vec{u}_L\left(-\frac{3P_L}{v_g^2}\frac{\vec{u}_L}{\tau_L}-\vec{\nabla}\check{\Pi}_E+n_L e \vec{E}\right)\nonumber\\
=&-\frac{3P_L}{v_g^2}\frac{\vec{u}_L^2}{\tau_L}-\vec{\nabla}\delta P \vec{u}_L+\vec{u}_L\vec{\nabla}\check{\sigma}-e\vec{j}\vec{\nabla}\phi.\nonumber\\
&=-\frac{3P_L}{v_g^2}\frac{\vec{u}_L^2}{\tau_L}-\frac{\partial u_{L,i}}{\partial x_j}\sigma_{ij}+\vec{\nabla}\left(\vec{u}_L\check{\sigma}-e\vec{j}\phi-\vec{u}_L\delta P\right).
\end{align}
The term $e n_L \vec{u}_L\vec{E} = e \vec{j}\vec{E}$ is the Joule
heating. Using the divergence theorem we can divide this into a
boundary and a bulk term
\begin{align}
0&=\mathcal{A}=\mathcal{A}_\text{boundary}-\mathcal{A}_\text{bulk},\\
\mathcal{A}_\text{boundary}&= 4\int\mathrm{d}\vec{A}\left(\vec{u}_L\check{\sigma}-\vec{u}_L\delta P- e \vec{j}\phi\right),\\
\mathcal{A}_\text{bulk}&=4 \int \mathrm{d}V \left( \frac{3P_L}{v_g^2}\frac{\vec{u}_L^2}{\tau_L}+\frac{\partial u_{L,i}}{\partial x_j}\sigma_{ij}\right).
\end{align}
The boundary term includes the energy transmitted through the interface.

Since the current density is conserved at the interface, we can
immediately write down the corresponding equation in the neutral
graphene sample, where the Joule heating is given by $ e
\delta\vec{j}\vec{E}$. Then we find
\begin{align}
0&=\mathcal{A}=\mathcal{A}_\text{boundary}-\mathcal{A}_\text{bulk},\\
\mathcal{A}_\text{boundary}&= 4\int\mathrm{d}\vec{A}\left(\vec{u}\sigma-\vec{u} \delta P-e\delta\vec{j}\phi\right),\\
\mathcal{A}_\text{bulk}&= 4 \int \mathrm{d}V \left(\frac{3P}{v_g^2}\frac{\vec{u}^2}{\tau_\text{dis}}+\frac{\partial u_i}{\partial x_j}\sigma_{ij}-\delta P(\vec{\nabla}\cdot\vec{u})\right).
\end{align}
As stated above, under realistic experimental conditions the
non-equilibrium part of the pressure at $u=0$ on the lead side
vanishes
\begin{align}
 \delta P=0.
\end{align}
At the same time, in neutral graphene sample we find
\begin{align}
 \delta P=\left(\frac{9 T^2\delta T \zeta (3)}{\pi  v_g^2}+\frac{\pi  \mu_I T^2}{3 v_g^2}\right)=\frac{T^2}{v_g^2}\left(\frac{9 \delta T \zeta (3)}{\pi  }+\frac{\pi  \mu_I}{3 }\right).
\end{align}
Using the hydrodynamic equations, one may replace $\delta P$ by $[-3
  P_B \tau_\text{RE}/(2r)]\partial(ru_r)/\partial r$, thus determining
$\delta P$ for finite $\tau_\text{RE}$ without any additional boundary
conditions. The same goes for $\mu_I$. Thus we may use the viscous
part of the dissipation to find the difference in the electrochemical
potential across the interface.

In addition, we may include the contact resistance described by 
\begin{align}
I^2 R_c=\vec{I}^T \check{R}\vec{I},
\end{align}
where $\vec{I}$ includes charge and entropy current and $\check{R}$
includes the thermoelectric coefficients of the interface.

In the absence of magnetic field, both $u_\vartheta$ and
$\sigma_{r\vartheta}$ vanish. In the leads $\delta P=0$ and hence we
find the condition
\begin{align}
4\pi\left[r\left(u_r\sigma_{rr}\right)\right]_{r_1-\epsilon}-2I\phi(r_1-\epsilon)=4\pi\left[r\left(u_r\sigma_{rr}- u_r  \delta P \right)\right]_{r_1+\varepsilon}-2I\phi(r_1+\epsilon)-2\vec{I}^T \check{R}\vec{I}\nonumber\\
\Leftrightarrow \phi(r_1-\epsilon)-\phi(r_1+\epsilon)=\frac{2\pi}{I}\left\lbrace \left[r\left(u_r\sigma_{rr}\right)\right]_{r_1-\epsilon}-\left[r\left(u_r\sigma_{rr}- u_r  \delta P \right)\right]_{r_1+\varepsilon}\right\rbrace+I R_c
\end{align}
at the first interface and similarly for the second interface
\begin{align}
4\pi\left[r\left(u_r\sigma_{rr}- \delta P u_r\right)\right]_{r_2-\epsilon}-2I\phi(r_2-\epsilon)=4\pi\left[r\left(u_r\sigma_{rr} \right)\right]_{r_2+\varepsilon}-2I\phi(r_2+\epsilon)-2\vec{I}^T \check{R}\vec{I}\nonumber\\
\Leftrightarrow \phi(r_2-\epsilon)-\phi(r_2+\epsilon)=-\frac{2\pi}{I}\left\lbrace \left[r\left(u_r\sigma_{rr}\right)\right]_{r_2+\epsilon}-\left[r\left(u_r\sigma_{rr}- u_r  \delta P \right)\right]_{r_2-\varepsilon}\right\rbrace+I R_c.
\end{align}

Combining the above general solution with these conditions, we find at $r_1$ 
\begin{align}
&r u_r(\sigma_{rr}-\delta P)=\frac{I^2 s_L^2 \eta\ell_\text{GE}}{4 \pi ^2 e^2  n_L^2 s_B^2\ell_G^2}\frac{ \ell_\text{GE}-r_2 K_0\left(\frac{r_1}{\ell_\text{GE}}\right) I_1\left(\frac{r_2}{\ell_\text{GE}}\right)-r_2 I_0\left(\frac{r_1}{\ell_\text{GE}}\right) K_1\left(\frac{r_2}{\ell_\text{GE}}\right)}{ r_1 r_2  \left[K_1\left(\frac{r_1}{\ell_\text{GE}}\right) I_1\left(\frac{r_2}{\ell_\text{GE}}\right)-I_1\left(\frac{r_1}{\ell_\text{GE}}\right) K_1\left(\frac{r_2}{\ell_\text{GE}}\right)\right]}-\frac{\eta  I^2 s_L^2}{2 \pi ^2 e^2 n_L^2 r_1^2 s_B^2}\nonumber
\end{align}
and at $r_2$
\begin{align}
&r u_r(\sigma_{rr}-\delta P)=-\frac{I^2 s_L^2 \eta\ell_\text{GE}}{4 \pi ^2 e^2 \ell_G^2 n_L^2 s_B^2}\frac{ \ell_\text{GE}-r_1 I_1\left(\frac{r_1}{\ell_\text{GE}}\right) K_0\left(\frac{r_2}{\ell_\text{GE}}\right)-r_1 K_1\left(\frac{r_1}{\ell_\text{GE}}\right) I_0\left(\frac{r_2}{\ell_\text{GE}}\right)}{r_1 r_2  \left[K_1\left(\frac{r_1}{\ell_\text{GE}}\right) I_1\left(\frac{r_2}{\ell_\text{GE}}\right)-I_1\left(\frac{r_1}{\ell_\text{GE}}\right) K_1\left(\frac{r_2}{\ell_\text{GE}}\right)\right]}-\frac{\eta  I^2 s_L^2}{2 \pi ^2 e^2 n_L^2 r_2^2 s_B^2}.\nonumber
\end{align}

As a result, we find the total resistance $R$ of the system in the form
\begin{align}
&I R=\phi(r_\text{in})-\phi(r_\text{out})=I (R_L+R_B+2 R_C+R_L^\text{diss}+R_B^\text{diss}),\\
&R_L =\frac{3 P_L }{2\pi e^2 n_L^2 v_g^2\tau_L }\log\left(\frac{r_1 r_\text{out}}{r_\text{in}r_2}\right),\quad R_B=\frac{R_0}{2\pi}\log\left(\frac{r_2}{r_1}\right),\\
&R_C =\frac{\vec{I}^T \check{R}\vec{I}}{I^2},\\
&R_L^\text{diss}=\frac{\eta_L}{\pi  e^2 n_L^2   }\left(\frac{1}{r_2^2}-\frac{1}{ r_1^2}\right)\\
&R_B^\text{diss}=\frac{\eta   s_L^2}{\pi e^2 n_L^2s_B^2}\left\lbrace \frac{1}{ r_1^2 }-\frac{1}{r_2^2 } +\frac{\ell_\text{GE}}{2 \ell_G^2  } \right.\\
&\times\left.\frac{  r_2\!\left[K_0\left(\frac{r_1}{\ell_\text{GE}}\right) I_1\left(\frac{r_2}{\ell_\text{GE}}\right)\!+\! I_0\left(\frac{r_1}{\ell_\text{GE}}\right) K_1\left(\frac{r_2}{\ell_\text{GE}}\right)\right] \!+\!r_1\!\left[ I_1\left(\frac{r_1}{\ell_\text{GE}}\right) K_0\left(\frac{r_2}{\ell_\text{GE}}\right)\!+\! K_1\left(\frac{r_1}{\ell_\text{GE}}\right) I_0\left(\frac{r_2}{\ell_\text{GE}}\right)\right]\!-\!2 \ell_\text{GE}}{r_1 r_2  \left(K_1\left(\frac{r_1}{\ell_\text{GE}}\right) I_1\left(\frac{r_2}{\ell_\text{GE}}\right)-I_1\left(\frac{r_1}{\ell_\text{GE}}\right) K_1\left(\frac{r_2}{\ell_\text{GE}}\right)\right)}\right\rbrace.\nonumber
\end{align}

\subsection{Analysis of results \label{sec:analysis}}

The behavior of the obtained resistance depends on the hierarchy of
length scales $r_1$, $r_2$, $r_2-r_1$, $\ell_\text{GE}$ and
$\ell_R$. In this Section, we specify the quantitative values of the
parameters used to produce the plots shown in the main text. For
clarity, here we restore the constants $\hbar$ and $k_B$.

We perform our quantitative analysis assuming the carrier density in
the leads to be $n_L=5\times 10^{14}$ m$^{-2}$. The equilibrium
temperature in the device (including both leads and the sample) is
fixed to $T=100$ K. The current, that is supplied by the source is
$I=1$ $\mu$A and we assume that the effective interaction constant is
screened to $\alpha=0.2$. We further use $\tau_\text{dis}=1.25\times
10^{-12}$ s and $\tau_L=0.189\times 10^{-12}$ s \cite{Kamada2021b},
since the density is higher in the leads. This determines all other
parameters, except for $\tau_\text{RE}$ and $\tau_R$ (or alternatively
$\ell_\text{GE}$ and $\ell_R$). Since these quantities are difficult
to extract from the available experimental data, we show results
for several different regimes.

The time scales related to electron-electron interaction are given by
\cite{Narozhny2021}
\begin{align}
\tau_{ii}= \hbar \frac{4\pi t_{ii}\log 2 }{\alpha^2 k_B T},
\qquad
t_{11}=\frac{1}{33.13},\quad t_{22}=\frac{1}{5.45}.
\end{align}
For the above parameter values, we find $\tau_{11}=0.5\times 10^{-12}$
s and $\tau_{22}=3\times 10^{-12}$ s. The viscosity can be estimated as
\begin{align}
\eta = \frac{0.446 k_B^2 T^2}{\alpha^2 v_g^2\hbar}
\end{align}
and amounts to $\nu=v_g^2\eta/(3 P)=0.25$ m$^2$/s. In addition
\begin{align}
R_0=\frac{\pi}{2\log 2}\frac{\hbar^2}{ e^2 k_B T}
\left(\frac{1}{\tau_{11}}+\frac{1}{\tau_\text{dis}}\right)=1985.33 \Omega
\end{align}

When describing the hydrodynamic velocity $u_r$ and the pressure $\delta P$ one can consider three different limits. If $\ell_\text{GE}\ll r_1,r_2$, which is achieved for very small $\tau_\text{dis}$, one finds
\begin{align}
u_r&\approx \frac{I s_L  \left(\sqrt{r r_2} \sinh \left(\frac{r-r_2}{\ell_\text{GE}}\right)-\sqrt{r r_1} \sinh \left(\frac{r-r_1}{\ell_\text{GE}}\right)\right)}{2 \pi  e n_L r s_B \sqrt{r_1 r_2} \text{sinh}\left(\frac{r_1-r_2}{\ell_\text{GE}}\right)}
\end{align}
which means, that the velocity vanishes exponentially close to the interface and is very small in the bulk of the sample. In the opposite limit $\ell_\text{GE}\gg r_1,r_2$ $u_r$ shows a behavior similar to the drift velocity in the leads with logarithmic corrections
\begin{align}
u_r&\approx \frac{I s_L   }{2 \pi  e  n_L r s_B }+\frac{I s_Lr_1^2 r_2^2 \log \left(\frac{r_1}{r_2}\right)}{4 \pi  e \ell_\text{GE}^2 n_L r s_B (r_1^2-r_2^2)}+\frac{I r s_L \left[r_1^2\log\left(\frac{r}{r_1}\right)-r_2^2\log\left(\frac{r}{r_2}\right)\right]}{4 \pi  e \ell_\text{GE}^2 n_L s_B \left(r_1^2-r_2^2\right)}.
\end{align}
Finally, if $r_2-r_1\ll r_1,r_2,\ell_\text{GE}$ we find the same $1/r$ behavior as in the leads
\begin{align}
u_r&\approx \frac{I s_L   }{2 \pi  e  n_L r s_B }
\end{align}
The resulting velocity $u_r$  is shown in Fig. 2 of the main text. In the leads, the drift velocity shows a simple $1/r$ behavior, while one finds a jump due to the mismatch of entropy directly at the interface. Inside the sample, the situation depends on the relative size of $\ell_\text{GE}$. If $\ell_\text{GE}\ll r_1,r_2$ we indeed observe, that the velocity decreases rapidly close to the interface and exactly vanishes in the bulk of the sample. This behavior is generally only observable in rather large samples, since the quantity $\tau_\text{dis}$ cannot be arbitrarily small while still staying in the hydrodynamic regime. In all other cases, $u_r$ resembles a $1/r$ behavior, that is slightly modified by logarithmic corrections.

The plot of $\delta T$ are shown in Fig. 3 of the main text. In the limit of $\ell_\text{GE}\ll r_1,r_2$ the non-equilibrium part of the temperature $\delta T$ vanishes in the bulk of the sample. In this limit energy relaxation processes transfer any heating, that may develop in the sample to the substrate and out of the device. There is only a small finite effect very close to the interface. Since this is an effect of $\tau_R$ it is in principle independent of $\ell_\text{GE}$ and $\tau_\text{RE}$, however we need $\ell_R< \ell_\text{GE}$ to remain in the hydrodynamic regime. In all other scenarios, there is a finite temperature profile, which may amount to $0.5\%$ of the equilibrium temperature.

Finally we take a look at the total resistance $R$ of the system. In general one might place the measuring points $r_\text{in}$ and $r_\text{out}$ very close to the interface, in which case the bulk resistance of the leads $R_L$ would not contribute to the total resistance $R$. We will further disregard the influence of the phenomenological contact resistance $R_C$, which only depends on the used materials and their relative chemical potential. Then one can consider again three limiting cases of the hydrodynamic, dissipative contribution to the resistance $R_B^\text{diss}$. The first limit is $\ell_\text{GE}\ll r_1,r_2$ in which case we find
\begin{align}
R_B^\text{diss}&\approx \frac{\eta   s_L^2}{\pi e^2 n_L^2s_B^2}\left(\frac{1}{ r_1^2 }-\frac{1}{r_2^2 } \right)\nonumber\\
&-\frac{ s_L^2 (A+\eta ) \left((r_1+r_2) \cosh \left(\frac{r_1-r_2}{\ell_\text{GE}}\right)-2 \sqrt{r_1 r_2}\right) \text{csch}\left(\frac{r_1-r_2}{\ell_\text{GE}}\right)}{2\pi e^2 \ell_\text{GE} n_L^2 r_1 r_2 s_B^2}\\
&\approx \frac{\eta   s_L^2}{\pi e^2 n_L^2s_B^2}\left[\frac{1}{ r_1^2 }-\frac{1}{r_2^2 }-\frac{\sqrt{\ell_G^2+\frac{v_g^2\tau_\text{dis}\tau_\text{RE}}{2}}}{2\ell_G^2}\left(\frac{1}{r_1}+\frac{1}{r_2}\right)\right]
\end{align}
where the second approximation requires $r_1-r_2\gg \ell_\text{GE}$.
The result of Ref. \cite{Shavit2019} corresponds to neglecting the term proportional to $\ell_\text{GE}$. The second limit is the case $\ell_\text{GE}\gg r_1,r_2$ and we find
\begin{align}
R_B^\text{diss}&\approx\frac{\eta   s_L^2}{\pi e^2 n_L^2s_B^2}\left(\frac{1}{ r_1^2 }-\frac{1}{r_2^2 } +\frac{1}{2 \ell_G^2}\log \left(\frac{r_2}{r_1}\right)\right),
\end{align}
 which introduces a logarithmic correction of exactly the same form as the bulk resistance $R_B$ of the sample. The final limit is $r_2-r_1\ll r_1,r_2,\ell_\text{GE}$ where we find the result
 \begin{align}
R_B^\text{diss}= \frac{ s_L^2 \eta   (r_2^2-r_1^2) }{4\pi e^2  n_L^2 r_1 r_2 s_B^2\ell_G^2}.
\end{align}
If one instead directly takes the limit $\tau_\text{RE}\rightarrow 0$, and additionally 
$\ell_G\ll r_1,r_2,r_2-r_1$ one would obtain
\begin{align}
R_B^\text{diss}\approx \frac{\eta   s_L^2}{\pi e^2 n_L^2s_B^2}\left[\frac{1}{ r_1^2 }-\frac{1}{r_2^2 }  -\frac{1}{2  \ell_G}\left(\frac{1}{r_1}+\frac{1}{r_2}\right)\right].
\end{align}
This is the result for the viscous correction to the resistance at charge neutrality in the setup of Ref. \cite{Li2022}.

The plots for $\phi(r)$ and $R=R_B+R_B^\text{diss}+R_L^\text{diss}$ are shown in Fig. 4 and 5 of the main text respectively. In the case of the potential $\phi$ we find a logarithmic dependence on the radial position $r$ in both the leads and the sample, where the overall prefactor is however different. In all considered cases, the jump at the interface is in the same direction, which for the second interface is opposite to what Ref. \cite{Shavit2019} obtains. This is due to the fact, that in our case the contribution of $\delta P$ is larger than the contributions due to $\eta$ and $\eta_L$ alone. The jump is larger, for larger $\ell_\text{GE}$. As seen in Fig. 5 of the main text, the total measured resistance is only slightly changed. The correction shown in the inset of Fig. 5 of the main text is nearly logarithmic for the larger $\ell_\text{GE}$, while is saturates for the smaller $\ell_\text{GE}$.

\bibliography{hydro-refs,refs-books, SuppRefs}
\end{document}